\documentclass[aps,twocolumn,superscriptaddress,showpacs]{revtex4-1}
\usepackage{amsmath,amssymb,multirow}
\usepackage{graphicx}
\usepackage{MnSymbol}
\begin{document}
\title{Strongly nonlinear thermovoltage and heat dissipation in interacting quantum dots}
\author{Miguel A. Sierra}
\affiliation{Instituto de F\'{\i}sica Interdisciplinar y Sistemas Complejos
IFISC (UIB-CSIC), E-07122 Palma de Mallorca, Spain}
\author{David S\'anchez}
\affiliation{Instituto de F\'{\i}sica Interdisciplinar y Sistemas Complejos
IFISC (UIB-CSIC), E-07122 Palma de Mallorca, Spain}
%
\begin{abstract}
We investigate the nonlinear regime of charge and energy transport through
Coulomb-blockaded quantum dots. We discuss crossed effects that arise
when electrons move in response to thermal gradients (Seebeck effect) or energy flows
in reaction to voltage differences (Peltier effect). We find that the differential thermoconductance
shows a characteristic Coulomb butterfly structure due to charging effects.
Importantly, we show that experimentally observed thermovoltage zeros 
are caused by the activation of Coulomb resonances at large thermal shifts.
Furthermore, the power dissipation asymmetry between the two attached electrodes
can be manipulated with the applied voltage, which has implications for the
efficient design of nanoscale coolers.
\end{abstract}
\maketitle
\textit{Introduction}.---In 1993 Staring {\it et al.} \cite{sta93} reported an intriguing
behavior of the thermovoltage $V_{th}$ generated across a thermally-driven
Coulomb-blockaded quantum dot. Their observations first indicated an increase of $V_{th}$ with
the temperature bias, in agreement with the Seebeck effect. Strikingly enough, for larger heating $V_{th}$
decreased, then vanished for a nonzero thermal difference and finally changed its sign.
Very recently, Fahlvik Svensson {\it et al.} \cite{fah13}
investigated the nonlinear thermovoltage properties of nanowires and made a similar observation.
The effect was attributed to a temperature-induced level renormalization because the piled-up
charge depends on the applied thermal gradient \cite{san13}. However, the potential response
was treated as a fitting parameter and single-electron tunneling processes were not properly taken
into account.

The subject is interesting for several reasons. First, Coulomb-blockade effects are ubiquitous
and govern the transport properties of a large variety of systems: quantum dots \cite{kou97},
molecular bridges \cite{par02}, carbon nanotubes \cite{pos01}, optical lattices \cite{chei08}, etc.
On the other hand, nanostructures are ideal candidates to test
novel thermoelectric effects boosting heat-to-work conversion performances \cite{dub11,ben13}.
Importantly, nonlinearities and rectification
mechanisms that lead to the phenomena reported in Refs.~\onlinecite{sta93,fah13} 
can be more easily tested in small conductors with strongly energy dependent densities
of states \cite{san13,boe01,kra07,kuo10,whi13,mea13,lop13,her13,mat13,hwa13,dut13,whi14,zim14,aze14}.
We emphasize that there is a close relation between the thermopower of a junction and its
heat dissipation properties, as demonstrated in Refs.~\onlinecite{lei10,lee13,ent14}
for the linear regime of transport. Therefore, ascertaining the conditions under which thermovoltages
acquire a significant nonlinear contribution has broader implications for power generation and
cooling applications \cite{gia06}.

We begin our discussion by noticing that vanishing thermovoltages imply the existence
of zero thermocurrent states. Unlike voltage-driven currents, which have a definite sign
for a bias voltage $V>0$ and never cross the $V$ axis for normal conductors
(an exception is the Hall resistance of an illuminated two-dimensional electron gas \cite{man02}),
electric transport subjected a thermal gradient $\theta$ displays regions of positive or
negative thermocurrents depending on the thermopower sign (positive for electron-like carriers,
negative for hole-like ones \cite{red07}). Nevertheless, this is not sufficient for the thermocurrent
to cross the $\theta$
axis since the thermopower is constant in linear response. Therefore, a strongly negative
differential thermoconductance $L=dI/d\theta$ is needed to drive the current $I$ from
positive to negative values. This results in an interesting effect---further
contact heating may switch off the thermocurrent flowing across the dot.
Notably, this is a purely nonlinear thermoelectric
effect and has no counterpart with either the voltage-driven case or the linear thermoelectric regime.

\textit{Theoretical model}.---Our results are based on the Anderson model
with constant charging energy $U$,
\begin{eqnarray}\label{eq_H}
\mathcal{H}=\mathcal{H}_{\rm leads}+\mathcal{H}_{\rm dot}+\mathcal{H}_{\rm tun}\,,
\end{eqnarray}
where $\mathcal{H}_{\rm leads}=\sum_{\alpha k \sigma} \varepsilon_{\alpha k \sigma} C^\dagger_{\alpha k \sigma} C_{\alpha k \sigma}$
is the Hamiltonian of left ($\alpha=L$) and right ($\alpha=R$) reservoirs coupled to the dot.
These are described as an electronic band of states with continuous wavenumber $k$ and spin index $\sigma=\{\uparrow,\downarrow\}$.
$\mathcal{H}_{\rm dot}=\sum_\sigma \varepsilon_d d_\sigma^\dagger d_\sigma+U d_\uparrow^\dagger d_\uparrow d_\downarrow^\dagger d_\downarrow$ is the dot Hamiltonian with quasilocalized level $\varepsilon_d$
(we consider a single level for definiteness).
$\mathcal{H}_{\rm tun}=\sum_{\alpha k \sigma}(V_{\alpha k \sigma}C_{\alpha k \sigma}^\dagger d_\sigma + \rm{h. c.})$
is the coupling term that hybridizes
dot and leads' states with tunneling amplitudes $V_{\alpha k \sigma}$.

The electronic current is given by the time evolution of the expected occupation
in one of the reservoirs, $I_\alpha=-e d \langle n_\alpha\rangle/dt$, with
$n_\alpha=\sum_{k \sigma} C^\dagger_{\alpha k \sigma} C_{\alpha k \sigma}$.
Since the total density commutes with the Hamiltonian of Eq.~\eqref{eq_H},
current conservation demands that $I_L+I_R=0$ in the steady state. Hence,
we can define the current flowing through the system as $I\equiv I_L=-I_R$.
Within the Keldysh formalism \cite{jauho}, $I$ is expressed as
$I= (e/\pi\hbar) \text{Re}\,\sum_{k\sigma}\int_{-\infty}^\infty dE \, V_{\alpha k \sigma} G_{\sigma , \alpha k \sigma}^<(E)$,
where $G_{\sigma , \alpha k \sigma}^<(E)=(1/\hbar)\int dE\, G_{\sigma , \alpha k \sigma}^<(t,t')e^{iE(t-t')/\hbar}$
is the Fourier transform of the lesser Green function
$G_{\sigma , \alpha k \sigma}^<(t,t')=\frac{i}{\hbar}\langle C_{\alpha k \sigma}^\dagger(t') d_\sigma(t)\rangle$.
Following Ref.~\cite{mei92}, the current readily becomes
\begin{equation}\label{eq_I2}
I =-\frac{e}{\pi\hbar} \int dE \sum_\sigma \frac{\Gamma_L\Gamma_R}{\Gamma}\text{Im}\,
G^r_{\sigma , \sigma}(E)[f_L(E)-f_R(E)]\,.
\end{equation}
$G^r$ is the dot retarded Green function in the presence of both coupling to the continuum states
and electron-electron interactions. $\Gamma_\alpha(E)=2\pi \rho_\alpha(E)|V_{\alpha \sigma}|^2$ denotes the level
broadening due to coupling to the leads (total linewidth $\Gamma=\Gamma_L+\Gamma_R$),
with $\rho_\alpha=\sum_k \delta(E-\varepsilon_{\alpha k})$
the $\alpha$ lead density of states. We consider the wide band
limit and take $\Gamma_\alpha$ as constant. Finally, in Eq.~\eqref{eq_I2} 
$f_\alpha(E)=1/[1+\exp{(E-\mu_\alpha)/(k_BT_\alpha)}]$ is the Fermi-Dirac function
for lead $\alpha$ with electrochemical potential $\mu_\alpha=E_F+eV_\alpha$
and temperature shift $T_\alpha=T+\theta_\alpha$ ($E_F$ is the common Fermi energy
and $T$ is the background temperature).

The spectral function given by $(-1/\pi)\text{Im}\,G^r$ in Eq.~\eqref{eq_I2} 
can be determined from the equation-of-motion technique followed by
a decoupling procedure \cite{hew62}. We restrict ourselves to the Coulomb blockade regime
($k_B T,\Gamma\ll U$)
and neglect cotunneling and Kondo correlations. This approach yields an excellent
characterization of the transport properties of strongly interacting quantum dots
for temperatures larger than the Kondo temperature, $T>T_K$. The retarded Green function
can be assessed by neglecting the correlators
$\llangle d_{\bar{\sigma}}^\dagger C_{\alpha k \bar{\sigma}} d_\sigma, d_\sigma^\dagger\rrangle \simeq 0$
and 
$\llangle  C^\dagger_{\alpha k \bar{\sigma}} d_{\bar{\sigma}} d_\sigma, d_\sigma^\dagger\rrangle \simeq 0$
(virtual charge excitations in the dot)
and
$\llangle C_{\alpha k \sigma} C_{\beta q \bar{\sigma}}^\dagger d_{\bar{\sigma}},d_\sigma^\dagger \rrangle \simeq 0$
and
$\llangle C_{\alpha k \sigma}  d_{\bar{\sigma}}^\dagger C_{\beta q \bar{\sigma}}, d_\sigma^\dagger \rrangle \simeq 0$
(spin excitations in the leads). Thus,
$G_{\sigma , \sigma}^r(E)=(1-\langle n_{\bar{\sigma}}\rangle)/(E-\varepsilon_d+i\Gamma/2)
+\langle n_{\bar{\sigma}}\rangle/(E-\varepsilon_d-U+i\Gamma/2)$
depends on the dot occupation for reversed spin $\bar{\sigma}$,
$\langle n_\sigma \rangle= \frac{1}{2\pi i} \int dE G^<_{\sigma , \sigma}(E)$.
Thus, $G^r$ must be calculated in a self-consistent fashion.
Using the Keldysh equation $G^<=i[\Gamma_L f_L(E)+\Gamma_R f_R(E)]|G^r|^2$,
we close the system of equations. $G^r$ has two poles at $E=\varepsilon_d$ and $E=\varepsilon_d+U$
broadened by $\Gamma$ and weighted by $(1-\langle n_{\bar{\sigma}}\rangle)$
and $\langle n_{\bar{\sigma}}\rangle$, respectively.
Then, the generalized transmission $(\Gamma_L\Gamma_R/\Gamma)\text{Im}\,
G^r_{\sigma , \sigma}(E,\{V_\alpha\},\{\theta_\alpha\})$
depends, quite generally, on both voltage and temperature shifts, as the occupation does,
which is a fundamental difference with noninteracting models \cite{but90}.

We find the spin-dependent occupations
\begin{eqnarray}
\langle n_\sigma \rangle &=& A(1-\langle n_{\bar{\sigma}} \rangle)+B\langle n_{\bar{\sigma}} \rangle\,,\\
\langle n_{\bar{\sigma}} \rangle &=& A (1-\langle n_\sigma \rangle)+B\langle n_\sigma \rangle\,,
\end{eqnarray}
where $A$ and $B$ are specified below.
The Hamiltonian in Eq.~\eqref{eq_H} is invariant under spin rotations since no Zeeman splitting
is present in the system. Hence, the mean occupation in the dot
$\langle n \rangle=\langle n_\sigma \rangle+\langle n_{\bar{\sigma}}\rangle$ is simply given by
\begin{equation}\label{eq_nAB}
\langle n \rangle = \frac{2A}{1+A-B}\,,
\end{equation}
with $A=(1/2\pi) \int dE \,[\Gamma_L f_L(E)+\Gamma_R f_R(E)]/[(E-\varepsilon_d)^2+\frac{\Gamma^2}{4}]$
and $B=(1/2\pi) \int dE \,[\Gamma_L f_L(E)+\Gamma_R f_R(E)]/[(E-\varepsilon_d-U)^2+\frac{\Gamma^2}{4}]$.
At equilibrium, $\langle n_\sigma \rangle=\langle n \rangle/2$
ranges between 0 and 1 depending on the value of $\varepsilon_d$,
which can be tuned with an external gate potential. As is well known,
the dot occupation significantly changes when $\varepsilon_d$ crosses the spectral
function peaks located at $E=E_F$ and $E=E_F+U$ (degeneracy points). In between, the
charge is approximately quantized. We now investigate departures of this behavior
when the dot is driven out of equilibrium due to either voltage or thermal gradients.

\begin{figure*}
\centering
\includegraphics[width=0.9\textwidth,clip]{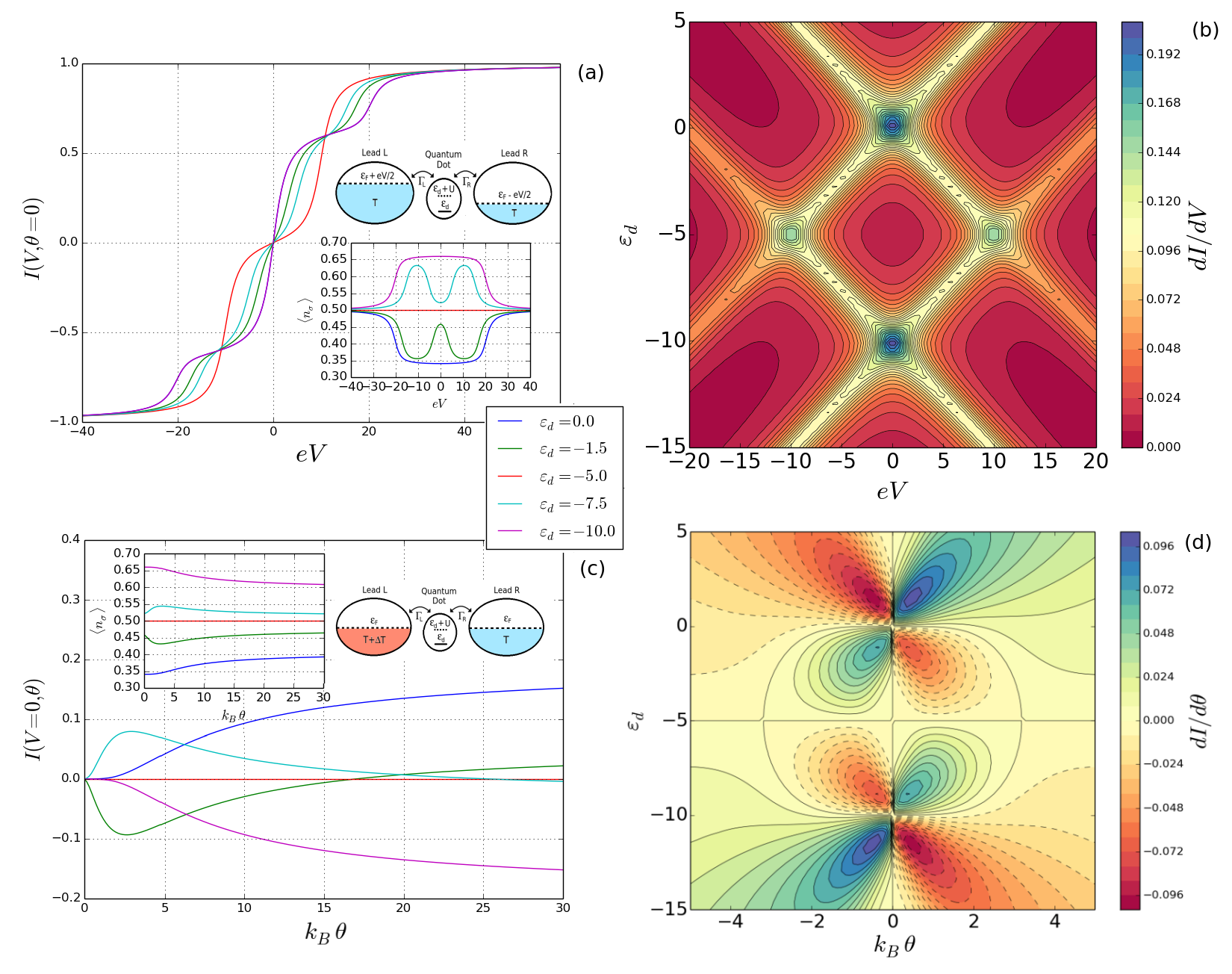}
\caption{(Color online). (a) Current--voltage characteristics of a dc-biased single-level
Coulomb-blockaded quantum dot (see the sketch) for the indicated gate voltages (level positions).
Inset: dot occupation as a function of the voltage bias. (b) Differential conductance
versus level position and bias voltage. (c) Thermocurrent of a single-level Coulomb-blockaded
quantum dot as a function of the temperature difference shown in the sketch.
(d) Differential thermoelectric conductance versus level position and bias voltage.
Parameters: charging energy $U=10$ and background temperature $k_B T=0.1$.
All energies are expressed in terms of $\Gamma_L=\Gamma_R=\Gamma/2$.
}\label{fig:1}
\end{figure*}

\textit{Voltage-driven case}.---We consider a voltage bias $V$ symmetrically
applied to the leads and set $E_F=0$ as the reference energy point, $\mu_L=-\mu_R=eV/2$.
Inserting Eq.~\eqref{eq_nAB} and the $G^r$ expression in Eq.~\eqref{eq_I2},
we calculate the $I$--$V$ characteristic curves for different values of the
dot level, see Fig.~\ref{fig:1}(a). When the single-particle peaks are at resonance with
the Fermi energy ($\varepsilon_d=0$ or $\varepsilon_d=U$), the system behaves as an ohmic junction
for voltages around $V=0$. With increasing $V$ the current reaches a plateau
and then increases again when the leads' electrochemical potential realigns with
the dot level, which causes an enhancement of the occupation as shown in the inset
of Fig.~\ref{fig:1}(a). This result \cite{mei91}
agrees with phenomenological models of Coulomb blockade~\cite{bee91}.
Clearly, the differential conductance $G=dI/dV$ traces show a Coulomb diamond structure as in
Fig.~\ref{fig:1}(b). 


The occupation is voltage independent in the particle-hole symmetry point
($\varepsilon_d=-U/2$), in which case the conductance is minimal around $V=0$. Only
for that case the transformation $d\to d^\dagger$ leaves Eq.~\eqref{eq_H} invariant and the electron
density in the dot follows a Fermi distribution. Away from $\varepsilon_d=-U/2$
the dot distribution is not Fermi-like since $A$ and $B$ become doubly stepped functions.
Therefore, the occupation [e.g., for $\varepsilon_d=-3U/4$ in the inset Fig.~\ref{fig:1}(a)]
exhibits a nonmonotonic dependence with $V$ and the conductance shows four peaks
as seen in Fig.~\ref{fig:1}(b).

\textit{Temperature-driven case}.---We
present in the bottom panel of Fig.~\ref{fig:1} the effect of a temperature shift $\Delta T>0$
applied to one of the electrodes: $\theta_L=\Delta T$ and $\theta_R=0$ for positive
temperature differences $\theta=T_L-T_R>0$, and $\theta_L=0$ and $\theta_R=\Delta T$
yielding $\theta<0$. Noticeably, the thermocurrent curves $I(\theta)$ in Fig.~\ref{fig:1}(c) lack the Coulomb staircases
seen in Fig.~\ref{fig:1}(a). For $\varepsilon_d=-U/2$ the thermocurrent is identically
zero since the dot spectral function of Eq.~\eqref{eq_I2} is symmetric around $E_F$.
At resonance, $I$ grows as the lead gets hotter because more thermally
excited electrons are able to tunnel through the nanostructure.
A similar response is obtained for level positions between $0$ and $U$ at small $\theta$.
Further increasing of $\theta$, however, gives rise to dramatic changes.
For $\varepsilon_d=-3U/4$ the thermocurrent reaches a maximum and then decreases, crossing the $\theta$
axis. In other words, a strong heating of one of the contacts reverses
the electronic flow, \textit{driving the electrons from the cold to the hot side}.
This striking behavior is opposite for gate potentials closer to the Fermi energy, see 
Fig.~\ref{fig:1}(c) for $\varepsilon_d=-0.15U$. This is a purely
nonlinear property of thermoelectric transport that is reflected
in the nonmonotonic occupation, see the inset of Fig.~\ref{fig:1}(c).

The differential thermoelectric conductance $L=dI/d\theta$
is shown in Fig.~\ref{fig:1}(d). The Coulomb diamonds of Fig.~\ref{fig:1}(b)
are transformed into a butterfly structure with strong changes
of sign across the points $\varepsilon_d=0$ and $\varepsilon=U$
for fixed $\theta$, in agreement with the experiment \cite{fah13}.
The effect is more intense for moderate values of the temperature shift $\theta\lesssim 10 T$,
a scale dominated by the charging energy.
As expected, we obtain $L=0$ for $\varepsilon_d=-U/2$ independently of $\theta$.
Above (below) this symmetry point, $L$
is positive (negative) in the small $\theta$ regime, which
is a manifestation of the Seebeck effect for electron-like
(hole-like) carrier transport. 

\begin{figure}
\centering
\includegraphics[width=0.45\textwidth,clip]{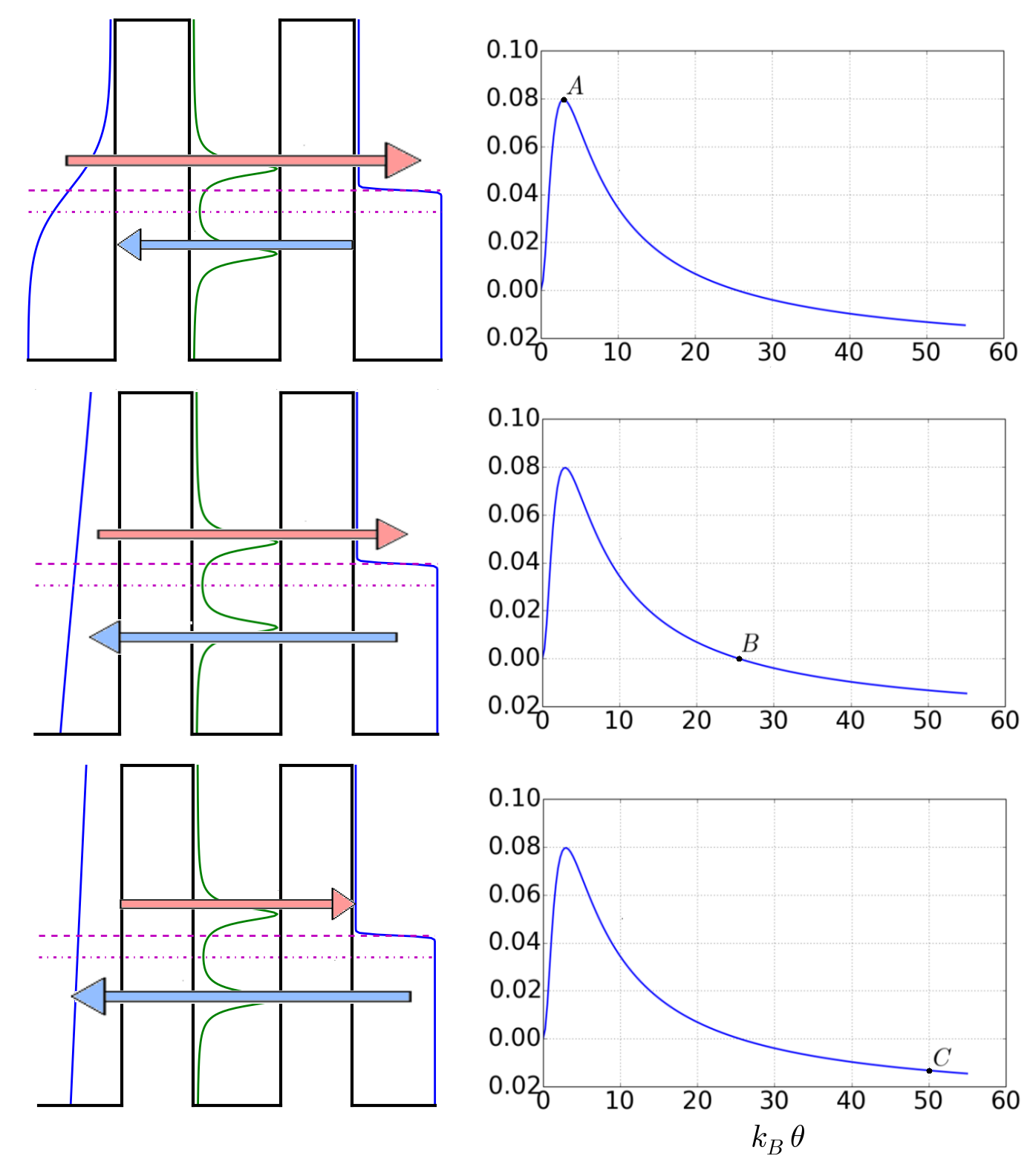}
\caption{(Color online). Left panel: energy diagram corresponding
to the current states of the right panel. $E_F$ ($\varepsilon_d$)
is indicated with dashed (dotted-dashed) lines. Right panel: thermocurrent
as a function of the temperature difference for $\varepsilon_d=-3U/4$
as taken from Fig.~\ref{fig:1}(c). Note that the electron flow
from the left (right) electrode at point $A$ ($C$) dominates but exactly cancels out
for point $B$.}\label{fig:2}
\end{figure}

\textit{Thermocurrent and thermovoltage}.---The strong
nonlinearities in the $I$--$\theta$ curves can be easily
understood with a level diagram as sketched in the left panel
of Fig.~\eqref{fig:2}. For $\varepsilon_d=-3U/4$ the $E=\varepsilon_d$
($E=\varepsilon_d+U$) pole lies below (above) $E_F$ (dot-dashed line).
If the left lead is heated, thermally excited electrons
contribute significantly to the current through the
$E=\varepsilon_d+U$ channel and the thermocurrent
becomes maximal [point labeled as $A$ in the right side
of Fig.~\eqref{fig:2}]. As the left contact becomes hotter,
the distribution function looses its step form unlike the
cold contact. As a consequence, holes (electrons traveling from the right
reservoir below $E_F$) counterbalance the flux from the left side,
giving rise to a vanishing thermocurrent at point $B$.
Further increase of $\theta$ causes a dominant contribution
of holes and $I$ takes on negative values (point $C$).

The thermovoltage or Seebeck voltage $V_{\rm th}$ is determined
from the open-circuit condition $I(V_{\rm th},\theta)=0$,
which we solve numerically to obtain $V_{\rm th}=V_{\rm th}(\theta)$.
Except for $\varepsilon_d=-U/2$, the thermovoltage
is generally nonzero. For a small thermal bias, $V_{\rm th}$
is a linear function of $\theta$, yielding a constant thermopower,
where the (differential) thermopower is defined as $S(\theta)=dV_{\rm th}/d\theta$.
For $\varepsilon_d$ close to $E_F+U$ ($E_F$), $S$ is positive (negative)
for $\theta\to 0$, which can distinguish transport due to electrons
or holes. With increasing $\theta$, the thermovoltage grows
because larger biases are needed to compensate the thermoelectric
flow. Hence, there exists a nice correlation between the $V_{\rm th}(\theta)$
and $I_{\rm th}(\theta)$ curves [cf. Fig.~\ref{fig:1}(c)].
For any value $\varepsilon_d\in (E_F,E_F+U)$ (except the special point
$\varepsilon=-U/2$) we always find a $\theta$ value such that
$V_{\rm th}=0$. The reason is clear from the above discussion.
For the point $B$ marked in Fig.~\eqref{fig:2} 
it is unnecessary to apply a voltage bias to counteract the thermal
gradient because the thermocurrent is already zero.
This effect would also be 
observable in dot systems with two levels but we remark that the
experiments of our interest \cite{sta93,fah13} are done in the Coulomb blockade regime.

\begin{figure}
\centering
\includegraphics[width=0.45\textwidth,clip]{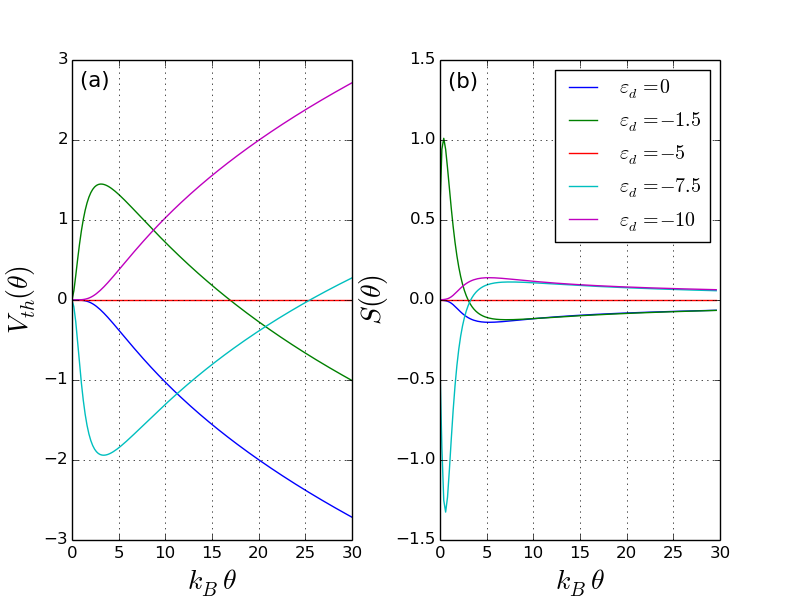}
\caption{(Color online). (a) Thermovoltage as a function
of the temperature difference for the indicated values
of the gate voltages (level positions). (b) Differential
thermopower $S=dV_{th}/d\theta$ in units of $k_B/e$. Parameters:
$U=10$, $k_B T=0.1$ and energy is given in units of
$\Gamma_L=\Gamma_R=\Gamma/2$.}\label{fig:3}
\end{figure}

\textit{Asymmetric dissipation and rectification}.---The reciprocal
effect to the Seebeck conversion is the Peltier effect \cite{kul94,bog99,zeb07}, which describes
a reversible heat that, unlike the Joule heating, can be used
to cool a system by electric means. Recent experiments \cite{lee13}
suggest an asymmetric rectification of the generated heat
in a voltage-driven atomic-scale junction.
These results are interesting because whereas rectification effects
are well understood in the electric case \cite{son98,lin00,sho01,fle02,but03,gon04,hac04}
much less is known about the way power is dissipated in a voltage-biased mesoscopic conductor.
The linear part of the rectified heat follows from the linear-response Peltier coefficient.
Therefore, the dissipated power can be larger or smaller for a given bias $V$
as compared with its reversed value, depending on whether the atomic resonance
lies above or below $E_F$.  However, nonlinear deviations were observed for larger $V$ \cite{lee13}.
Here, we demonstrate that the heat rectification can be tuned with $V$
for a \textit{fixed} position of $\varepsilon_d$.

\begin{figure}
\centering
\includegraphics[width=0.45\textwidth,clip]{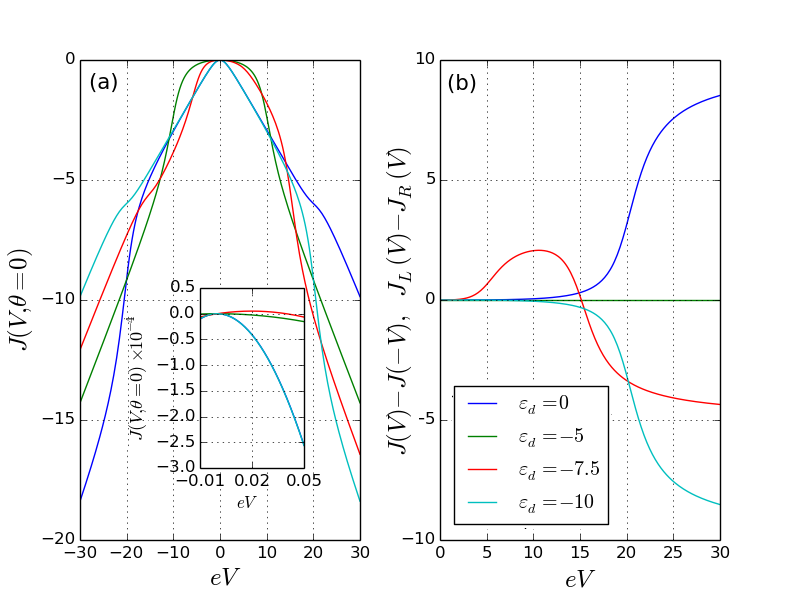}
\caption{(Color online). Heat current as a function of applied voltage
in the isothermal case $\theta=0$. Dot level positions
are also indicated. Inset: Detail of the dissipated power
around zero voltage. (b) Asymmetric dissipation versus voltage bias
for the same gate voltages. Parameters:
$U=10$, $k_B T=0.1$ and energy is given in units of
$\Gamma_L=\Gamma_R=\Gamma/2$.}\label{fig:4}
\end{figure}

The heat current is derived from $J_\alpha=d \langle \sum_{k \sigma} \varepsilon_{\alpha k \sigma} C^\dagger_{\alpha k \sigma} C_{\alpha k \sigma}\rangle/dt$,
\begin{eqnarray}\label{eq_J2}
J_\alpha =\sum_\sigma \frac{\Gamma_L\Gamma_R}{\pi\hbar\Gamma}\!\!\int\!\! dE \,(\mu_\alpha-E)\text{Im}\,
G^r_{\sigma , \sigma}[f_L(E)-f_R(E)]\,,
\end{eqnarray}
which satisfies the Joule law $J_L+J_R=-IV$. We consider the case where
$J\equiv J_L$ is a function of voltage only ($\theta=0$).
Figure~\ref{fig:4}(a) shows the heat current as a function of $V$ for
several values of $\varepsilon_d$. Only for $\varepsilon_d=-U/2$, 
$J$ exhibits a symmetric behavior, as expected. We observe that
the curves quickly depart from the linear regime [see the inset of Fig.~\ref{fig:4}(b)].
Thus, Joule and higher order effects start soon to dominate.
Interestingly, for $\varepsilon_d=-3U/4$ the heat current shows a nontrivial
zero for finite $V$.
The resulting asymmetry under $V$ reversal is apparent for, e.g., $\varepsilon_d=U$.
Our results also show that the heat current is invariant under the 
joint transformation $V\to -V$ and $\varepsilon_d\to -\varepsilon_d-U$
(see, e.g., the $\varepsilon_d=0$ and $\varepsilon_d=-U$ cases).

In Fig.~\ref{fig:4}(b) we depict the rectification factor $J(V)-J(-V)$
for different dot level positions. At resonance ($\varepsilon_d=0$),
the rectification is always positive, i.e., the dissipation is larger for $V>0$ than for
$V<0$ and increases with voltage. Clearly,
for $V>0$ heat can flow through either $E=0$
or $E=U$ peaks while for $V<0$ energy can be transported through the $E=0$
resonance only. The situation is reversed for $\varepsilon_d=-U$. The
dissipation is now larger for negative voltages than for positive polarities.
More importantly, the rectification factor can change its sign for
a given value of $\varepsilon_d$, as indicated in  Fig.~\ref{fig:4}(b)
for $\varepsilon_d=-3U/4$. Notice that this is a purely nonlinear
effect. While in the linear case the rectification can be changed
with tuning $\varepsilon_d$ and this effect heavily depends
on the transmission energy dependence \cite{lee13}, here a voltage
scan leads to a value where the transformation
$V\to -V$ leaves $J$ invariant. Furthermore, it is straightforward
to show that the power difference between the left and right electrodes for a given bias,
$J_L(V)-J_R(V)$, equals $J(V)-J(-V)$ if the dot spectral function is symmetric
under $V$ reversal. Our system indeed shows this property for symmetric couplings,
$\Gamma_L=\Gamma_R$ [see the inset of Fig.~\ref{fig:1}(a), where the occupation
is an even function of $V$].
Therefore, heat can be dissipated equally between the leads ($J_L=J_R$) for $V\neq 0$
[see Fig.~\ref{fig:4}(b) for $\varepsilon_d=-3U/4$],
despite the fact that the transmission strongly depends on energy, unlike the linear case  \cite{lee13}.
This is again an effect which can be observed in the nonlinear
regime of transport only.

\textit{Conclusion}.---We have examined a counterintuitive phenomenon
seen in experiments--with increasingly thermal gradient applied to a
quantum dot the created thermovoltage
diminishes and even becomes zero for a nonzero temperature bias.
We have shown that the effect is due to the combined influence
of the {\em two} peaks arising from a Coulomb blockade level. Furthermore,
we predict a reciprocal effect--the power rectification becomes zero
for a {\em finite} voltage, which can be relevant for the design
of nanodevices with controllable dissipation. Further work
should clarify the role of (higher-order) cotunneling processes \cite{tur02,tor13}
and Kondo interactions \cite{don02,sch05,cos10}.

\textit{Acknowledgments}.---We thank R. L\'opez for useful discussions.
Work supported by MINECO under Grant No.~FIS2011-23526 and MECD.

\end{document}